\pdfoutput=1
\documentclass[lefttitle]{elsarticle}

\usepackage{graphicx}
\usepackage{subfigure}
\usepackage{fullpage}
\usepackage{hyperref}
\usepackage{amsmath}
\DeclareMathOperator{\sgn}{sgn}

\usepackage{soul}
\usepackage{xspace}

\usepackage{algorithmic}
\usepackage{algorithm}
\floatname{algorithm}{Game}

\newcommand{\game}[1]{\textsf{#1}}
\newcommand{\gameA}{\game{A}\xspace}
\newcommand{\gameB}{\game{B}\xspace}

\newcommand{\figref}[1]{Fig.~(\ref{#1})}
\newcommand{\gameref}[1]{\textbf{Game~\ref{#1}}\xspace}
\newcommand{\eps}{\ensuremath{\epsilon}}

\date{02/10/2021 (v. 0.30)}

\journal{Physica A: Statistical Mechanics and its Applications}
\begin{document}

\begin{frontmatter}

\title{Constructing games on networks for controlling the inequalities in the capital distribution}
\author[1]{Jaros\l aw Adam Miszczak}
%\affiliation[1]{organization={Institute of Theoretical and Applied Informatics, Polish Academy
%		of Sciences},
%	addressline={Baltycka 5},
%	postcode={44-100},
%	city={Gliwcie},
%	country={Poland}}

\address{Institute of Theoretical and Applied Informatics, Polish Academy
of Sciences,\\ Ba{\l}tycka 5, 44-100 Gliwice, Poland}
\ead{jmiszczak@iitis.pl}
\ead[url]{https://orcid.org/0000-0001-8790-101X}

\begin{abstract}
The inequality in capital or resource distribution is among the important phenomena observed in populations. The sources of inequality and methods for controlling it are of practical interest. To study this phenomenon, we introduce a model of interaction between agents in the network designed for reducing the inequality in the distribution of capital. To achieve the effect of inequality reduction, we interpret the outcome of the elementary game played in the network such that the wining of the game is translated into the reduction of the inequality. We study different interpretations of the introduced scheme and their impact on the behaviour of agents in the terms of the capital distribution, and we provide examples based on the capital dependent Parrondo's paradox. The results presented in this study provide insight into the mechanics of the inequality formation in the society. 
\end{abstract}

\begin{keyword}
social inequality \sep Matthew effect \sep Parrondo's games \sep agent-based modeling
\MSC[2010] 05C57\sep  62C86
\end{keyword}

\end{frontmatter}

%\tableofcontents

%\linenumbers
%%%%%%%%%%%%%%%%%%%%%%%%%%%%%%%%%%%%%%%%%%%%%%%%%%%%%%%%%%%%%%%%%%%%%%%%%%%%%%%%
\section{Introduction}
%%%%%%%%%%%%%%%%%%%%%%%%%%%%%%%%%%%%%%%%%%%%%%%%%%%%%%%%%%%%%%%%%%%%%%%%%%%%%%%%
The inequality in the communities has been studied by various authors and numerous methods for describing and measuring inequalities have been proposed~\cite{cowell2011measuring,milanovic2011worlds}. However, it is not clear why this phenomenon emerges in communities, and how to mitigate it. As the qualitative analysis of the populations resembles in many aspects the analysis of interactions between agents or particles, the natural approach for modelling the dynamics of inequalities is to use the methods borrowed from statistical physics. The field of socio-physics aims at applying the rules of statistical mechanics for exploring the phenomena observed in social structures, including the human society~\cite{castellano2009statistical}.

One of the mechanisms appearing in many scenarios analysing the global behaviour of different populations is the effect of cumulative advantage.  The mechanism of cumulative advantage has also been observed in the various aspects of human society~\cite{perc2014matthew}. This interesting effect is observed in many small communities, with the community of scientists being the first one considered. In particular, in the sociology of science, the term Matthew effect has been coined~\cite{merton1968matthew} to refer to the effect of eminent scientists getting more credit than a less known researchers, even if their contribution is similar.  Matthew effect is obeyed by small-world networks, where it is observed as the preferential attachment mechanism. In the context of network formation this mechanism can be used to explain the formation of World Wide Web  network~\cite{price1965networks, price1976general, barabasi1999emergence, price1976general} giving rise to the preferential attachment process.

Recently, Matthew effect has been studied in the context of sociological competition. In particular in \cite{karataieva2019mean} a model capturing the acquisitions of power -- understood as a total energy -- during the interaction between the agents has been proposed.  On the other hand, in \cite{hu2007simulating} the authors investigated a model of evolutionary game with rules evolving in each round so that the agent tends to copy the strategy of the most successful agent from their neighbourhood.

Even if many models of describing the rise of inequality have been proposed, it is equally interesting to ask to what degree one can control the rise of inequalities. For these reasons, in this paper we propose a simple model constructed for the purpose of mitigating the inequality of wealth distribution in the population. It can be argued that providing such mechanisms is important fom the purpose of social science.  It is also worth noting that inequalities and diversity in the population can be used to explain the emergence of cooperation~\cite{perc2008social,santos2008social,szolnoki2008diversity}. However, one should note that the inequality in the distribution of resources is not unique for the human society. On the contrary, a similar mechanisms can be used to explain inequality in the technical systems. In such case the agents ca represent software agents acting in the computer network or processes in the operating system.

The main contribution of the presented paper is the study of the mechanisms which can be used to mitigate inequality in the distributed system of interacting agents. We introduce a simple scheme in which a selected two-player game can be used for controlling the inequality of capital distribution in the population. To this end we interpret the winning not as the increase in the capital, but rather as the decrease in the inequality between the agents participating in the game. We show that the Matthew effect can be defined as  an example of the policy for interpreting the results of the game. To study the effects of such policies we analyse the behaviour of the inequalities using alternative polices.

As a special example of the method used for controlling the inequality of capital distribution we use the game constructed using capital dependant Parrondo's scheme. In particular, we study the interplay between the parameters of the Parrondo's scheme and the resulting inequality reduction. We argue that the appearance of the paradoxical behaviour does not have a straightforward interpretation in terms of the capital distribution control.

This paper is organized in the following manner. In Section~\ref{sec:prelimianaries} we introduced basic concepts used in this work. In Section~\ref{sec:janosik-model} we describe a framework of building games on networks suitable for controlling the inequality in the capital distribution. In Section~\ref{sec:policy-seclection} we provide examples of scenarios where the inequality of capital distribution is controlled in various manners. In particular, we study random policy selection and a policy based on Parrondo's scheme. Finally, in Section~\ref{sec:discussion} we summarize the presented results, provide a discussion of the presented results, and outline possible extensions and applications.

%%%%%%%%%%%%%%%%%%%%%%%%%%%%%%%%%%%%%%%%%%%%%%%%%%%%%%%%%%%%%%%%%%%%%%%%%%%%%%%%
\section{Preliminaries}\label{sec:prelimianaries}
%%%%%%%%%%%%%%%%%%%%%%%%%%%%%%%%%%%%%%%%%%%%%%%%%%%%%%%%%%%%%%%%%%%%%%%%%%%%%%%%

Let us start by introducing concepts and fixing the notation used in this work. We assume that our population consists of $n$ agents. The agents are free to interact which each of their neighbours and the neighbourhood is defined by the structure of the graph. Additionally, the agent can move according to the structure of the underlying graph.

Our goal is to analyse a simple scheme of mitigating the inequality in the capital distribution in the population. We will focus on the capital of money possesed by the individuals~\cite{dragulescu2000statistical}. 
The scheme discussed in this work assumes that capital obeys the conservation law and was inspired by the approach applied in \cite{dragulescu2000statistical,dragulescu2003statistical, yakovenko2009colloquium}. Moreover, for the sake of simplicity we do not consider debt. One can also note that the described model can be used to describe any kind of resources which can be exchanged by the individuals as long as they obey the conservation law and can be represented by non-negative numbers.

The amount of money possessed by individuals is one of the ingredients of the wealth and the distribution of money can be analysed using the statistical mechanics approach~\cite{dragulescu2000statistical,dragulescu2003statistical}. In the case where only the conservation law is assumed, it can be shown that in the equilibrium the distribution of money obeys Boltzmann-Gibbs law, with probability mass function
\begin{equation}
P(m_i) \sim \exp^{-m_i/T},
\end{equation}
where the effective temperature is given by the average amount of money per agent, $T=\frac{\sum_{i=1}^n m_i}{n}$. Here $m_i\in\{1, 2, \dots, M\}$, where $M$ is the total amount of money distributed among the agents. Such equilibrium distribution of money can be derived under the assumption that money obey the conservation law~\cite{dragulescu2000statistical}.

The main goal of the presented work is to analyse the inequalities in the capital distributions between the agents on the networks.
For the purpose of quantifying the effects of the introduced strategies for mitigating the inequalities we use commonly used metrics used to measure the inequality in the population. One of the most popular metrics used for the purpose of measuring the inequality of the capital distribution is \emph{Gini index}. This quantity is commonly adopted as a measure of statistical dispersion used to describe the economic inequality~\cite{glasser1962variance}. For the vectors of positive numbers $\textbf{x} = (x_1,x_2,\dots,x_n)$, Gini index is defined as an average of the absolute differences for all pairs of elements,  
\begin{equation}
\mathcal{G}(x) = \frac{\sum_{i=1}^n\sum_{j=1}^n |x_i-x_j|}{n\sum_{i=1}^n x_i}.
\end{equation}

The Gini index can range from zero, meaning complete equality in the population, to one, which indicates complete inequality. In what follows, we use the Gini index to measure the inequality of the capital in the population of agents on the network. The no-debt assumption is crucial in this case, as it is not possible to use the Gini directly index in the case of negative values.

\begin{figure}[ht!]
\centering
\subfigure[constant increase of the capital]{
\includegraphics[scale=0.95]{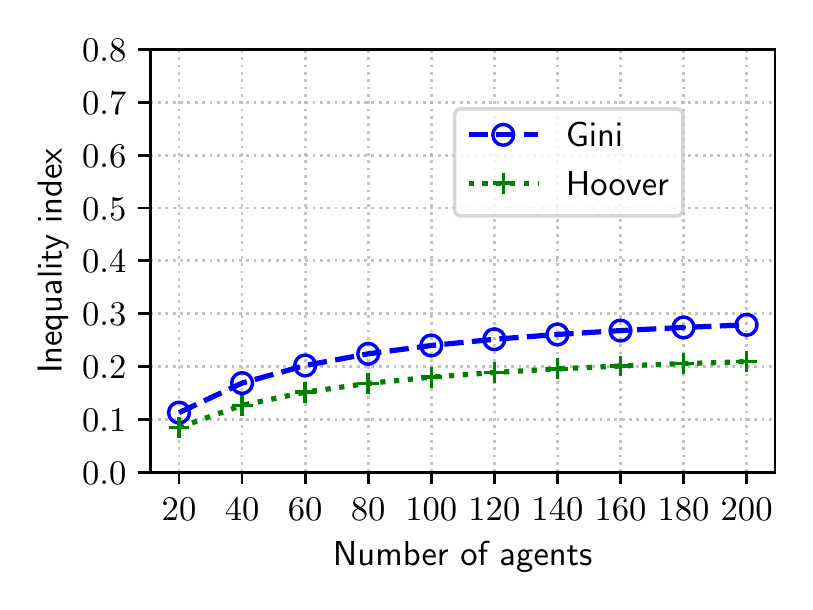}\label{fig:init-gini-hoover-constant}}%
\subfigure[equilibrium distribution of the capital]{
\includegraphics[scale=0.95]{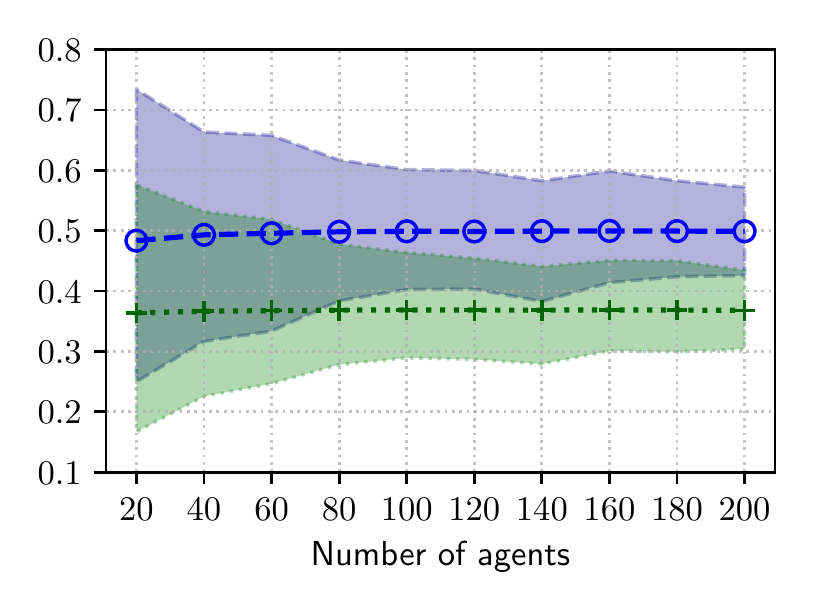}\label{fig:init-gini-hoover-bg}}

\caption{Initial value of the inequality indices -- Gini index and Hoover index -- obtained for the capital initialization \subref{fig:init-gini-hoover-constant} based on the constant increase of the capital give by Eq.~\eqref{eq:init-wealth-const}, and \subref{fig:init-gini-hoover-bg} equilibrium distribution given for Boltzmann-Gibbs distribution~from Eq.~\eqref{eq:bg-dist}. In the second case, $10^5$ initial samples were used for each number of agents $n=20,40,\dots,200$, and each sample represents a possible initial state used in the numerical experiments starting in the equilibrium.}
\label{fig:init-gini-hoover}
\end{figure}

One should note that a plethora of inequality measures is used in the literature~\cite{cowell2011measuring,milanovic2011worlds}. Another quantity commonly used to measure the inequality in the income distribution is Hoover index, also know as Robin Hood\footnote{Legendary heroic outlaw originally depicted in English folklore.} index. Hoover index is defined as the total amount of capital which have to be redistributed to ensure that every agent posses equal capital,
\begin{equation}
\mathcal{H}(x) = \frac{1}{2} \frac{\sum_{i=1}^n |x_i - \frac{1}{n}\sum_{i=1}^n x_i|}{\sum_{i=1}^n x_i}.
\end{equation}
As in the case of Gini index, Hoover index ranges from zero, in which case all agents posses the same amount of the capital, to one, which is the case of maximal inequality. Comparison og Gini and Hoover index for two different case of capital distribution is presented in \figref{fig:init-gini-hoover}. However, in this paper we focus on the Gini index. 

%Another metric used to measure economic inequality is provided by Theil index. This quantity is on the information theory. Theli index, which is an example of generalized entropy index, is defined as maximum possible entropy of the data minus the observed entropy,
%\begin{equation}
%\mathcal{T}(x) = \frac{1}{n}\sum_{i=1}^n \frac{x_i}{\frac{1}{n}\sum_{i=1}^n x_i} \ln \frac{x_i}{\frac{1}{n}\sum_{i=1}^n x_i}.
%\end{equation}

As we aim at analysing the scheme the inequality, we need to consider various initial distributions of the capital.
For the purpose of this work we will consider two scenarios for the initial capital distribution. In the first scenario the initial value of the wealth assigned to the agents was set to
\begin{equation}\label{eq:init-wealth-const}
m_i=W+i,
\end{equation}
where $i=1,2,\dots,n$ is the number of agents, and $W$ is the base value of the capital. Using these initialization values we get the initial behaviour of Gini index as depicted in \figref{fig:init-gini-hoover}.

In the second scenario, in order to check the behaviour of the model in the situation when we start from the equilibrium, we will use Boltzmann-Gibbs distribution of the initial capital. We assume that the initial amount of money is the same as in the previous initialization strategy. The probability mass function is given by
\begin{equation}\label{eq:bg-dist}
P(w) \sim \exp(-\frac{wn}{M}),
\end{equation}
for $w=0,1,\ldots,M$, where $M=2n+\frac{n^2}{2}$ and the effective temperature $T=2+\frac{n}{2}$. This leads to the minimum and maximum values of the Gini index as depicted in \figref{fig:init-gini-hoover-bg}.

One should note that Gini index tends to be small for small populations~\cite{deltas2003small-sample}. This is visible in the case of initialization with a constant increase of capital depicted in \figref{fig:init-gini-hoover-constant}. At the same time, in the case of Boltzmann-Gibbs distribution, for small populations one can observe larger difference between the minimal and the maximal value of the initial value of Gini coefficient (cf. \figref{fig:init-gini-hoover-bg}). For this reason, one can start from the smaller initial value of Gini index when a small population is considered.

As an particular example of the method of choosing the elementary games played by the agents in the network we consider the example of the Parrondo's effect applied in the proposed scheme. \emph{Parrondo's paradox} or \emph{Parrondo's game} is a scheme used to combine two games with negative gain in order to obtain a scheme with a positive payoff~\cite{harmer1999losing, harmer2002review}. The scheme can be used in various scenarios and the paradoxical effect has been observed in various setups~\cite{gawron2005quantum,kovsik2007quantum,pawela2013cooperative,lai2020parrondo}.
For the case of games on networks, which is most relevant for the purpose of this work, it was demonstrated in \cite{ye2011cooperation, ye2016effects} that the combination of two loosing games can lead to positive gain. 

For more details one can consult recent results on Parrondo's games on graphs and complex networks \cite{ye2011cooperation, ye2016effects, ye2021effects}. For a comprehensive review of the recent investigations concerning the application of the Parrondo's scheme in the context of social dynamics see \cite{lai2020social}.

In the context of this work we are interested in the connections between the policy selection based on Parrondo's scheme and the resulting inequality reduction. Thus, we utilize the scheme used to obtain the paradoxical effect observed by Parrondo, and we use it as the elementary process for controlling the redistribution of capital in the population. 

%%%%%%%%%%%%%%%%%%%%%%%%%%%%%%%%%%%%%%%%%%%%%%%%%%%%%%%%%%%%%%%%%%%%%%%%%%%%%%%%
\section{Network game for inequality control}\label{sec:janosik-model}
%%%%%%%%%%%%%%%%%%%%%%%%%%%%%%%%%%%%%%%%%%%%%%%%%%%%%%%%%%%%%%%%%%%%%%%%%%%%%%%%

Let us consider a population of $n$ agents on the network. Initially each of the agents has some amount of capital $m_i$, $i=1,2,\dots,n$. At each time-step an agent selects randomly one mate from the current node. Next, the agent plays a game against the selected mate. Finally, the result of the game is used to alter the capital of both players.

In the simplest cast, the game played by the agents is a coin-flip, which can be possibly biased. The modification of the capital is based on the result. In \gameref{proc:janosik-v1}, such scheme on a graph $G$ with $n$ agents and an initial wealth assigned to each agent equal to $m_i$ is presented. We assume that the coin flip game played by the agents can have a non-zero bias, $\eps\in[0,\frac{1}{2}]$, toward loosing. 

\begin{algorithm}[H]
\begin{algorithmic}
    \REQUIRE coin-flip game \gameA with game bias $\eps\in[-\frac{1}{2},0]$, graph $G$, $m_i \geq 0, i=1,2\dots, n$, 
    \FOR{$i=1$ to $n$}
        \STATE $j$ $\leftarrow$ another agent from the current node  such that $m_j>1$ \COMMENT{Avoid the debt}
        \STATE $r$ $\leftarrow$ result of the $\gameA$ played by agent $i$ \COMMENT{Positive outcome means win for agent $i$}
       	\STATE $\Delta \leftarrow \sgn{(m_i-m_j)}$
    \IF{$r > 0$} 
      \STATE $(m_i, m_j) \leftarrow (m_i-\Delta, m_j+\Delta)$ \COMMENT{Decrease the difference between $m_i$ and $m_j$}
    \ELSE
      \STATE $(m_i, m_j) \leftarrow (m_i+\Delta, m_j-\Delta)$ \COMMENT{Increase the difference between $m_i$ and $m_j$}
    \ENDIF
        \STATE {select one of the neighbour nodes for the current node occupied by $i$}
        \STATE {move agent $i$ to the new position}
    \ENDFOR
\end{algorithmic}
\caption{Scheme for the direct interpretation of the results of the coin-flip game $\gameA$ played on the network.}
\label{proc:janosik-v1}
\end{algorithm}

The crucial element distinguishing the proposed scheme from the standard method of defining the game on graph is the interpretation of the result. In the proposed case the difference in the value of capital $m_i$ is used in the course of the scheme. Instead of altering their capital according to the outcome of the game, agent interpret the outcome of the game in such manner that winning leads to the action reducing the inequality in the capital distribution. In particular, if agent $i$ wins against agent $j$, they both agree to increase the capital of the agent with the smaller capital, and reduce the capital of the agent with larger capital. On the other hand, if the agent loses, the difference is amplified. Thus, the scheme is constructed with the goal of controlling the distribution of capital in the population of agents.

We assume that agents are randomly distributed on the network. Moreover, each agent has ability to move to the nearest nodes. In the case considered in this work we limit our considerations to the case where the underlying network is a 2D grid with periodic boundary conditions. Preliminary simulation results suggest that the structure of the network has no clear impact of the reduction of inequalities. However, only basic experiments were performed.

Capital possessed by each agent is governed by two processes. At each step, each agent plays a game. Subsequently, the result of the game is interpreted in the process of capital update. In summary, each step consists of the following actions.
\begin{itemize}
\item If there are any other agents at the current node, play a selected game against a randomly selected opponent.
\item Using the outcome of the game, update the capital of both players such that, in the case of the win for the current player, the difference will decrease.
\item Choose one of the neighbouring nodes and move to a new location.
\end{itemize}

The pseudocode for the above scheme is presented in \gameref{proc:janosik-v1}.
One should note that the agent will play only if there are other agent at the node it currently occupies. Thus one should expect faster changes in the Gini index for the networks with more connections and for the situation when there are more agents in the network.

Moreover, one can consider a case where the model included additional bias parameters introduced for the consideration of \emph{unfair} interpretations of the results of the game \gameA. Such scheme is provided in \gameref{proc:janosik-v2}, where the additional parameters $[b_+,b_-]$ are considered in the process of updating the capital of players engaged in each elementary game. One should note that in this case we deal with the biased game played at each step and the biased interpretations of the results. However, we do not consider the probabilistic interpretation of the results and the selection of bias $[b_+,b_-]$ is made at the beginning of the game.

\begin{algorithm}[H]
\begin{algorithmic}
    \REQUIRE game \gameA, game bias $\eps\in[-\frac{1}{2},0]$, graph $G$, $m_i \geq 0, i=1,2\dots, n$, and policy bias $b=[b_{+},b_{-}]$
    \FOR{$i=1$ to $n$} 
        \STATE $j$ $\leftarrow$ choose another agent from the node occupied by agent $i$
        \STATE $r$ $\leftarrow$ result of the game $\gameA$ played by agent $i$
       	\STATE $\Delta \leftarrow \sgn{(m_i-m_j)}$
    \IF{$r > 0$} 
      \STATE $(m_i, m_j) \leftarrow (m_i-b_+\Delta, m_j+b_+\Delta)$
    \ELSE
      \STATE $(m_i, m_j) \leftarrow (m_i+b_-\Delta, m_j-b_-\Delta)$ 
    \ENDIF
        \STATE {Select one of the neighbour nodes for the current node occupied by $i$.}
        \STATE {Move agent $i$ to the new position.}
    \ENDFOR
\end{algorithmic}
\caption{Scheme for the interpretation of the results of game $\gameA$, with policy bias.}
\label{proc:janosik-v2}
\end{algorithm}

In comparison to the procedure \gameref{proc:janosik-v1}, procedure provided in \gameref{proc:janosik-v2} includes additional pair of parameters, $b$, which influences the interpretation of the wining and loosing. Thanks to this, the scheme defined in \gameref{proc:janosik-v2} can be used to control the interpretation of the winning in the game~\gameA used in the scheme. In particular, one can consider the following policies.
\begin{itemize}
\item \emph{Janosik\footnote{Janosik was a Slovak folk hero taking from the rich and giving to the poor, main character of many legends taking place in Tatra mountains on Polish-Slovak border.} policy} or \emph{anti-Matthew policy}, $[b_+, b_-] = [1,1]$, leading to the decrease of the difference of capital between agents if the game outcome is positive.  

\item \emph{Mathhew policy}, $[b_+, b_-] = [-1,-1]$, leading to the interpretation of the positive outcome as a increase in the difference between the agents.

\item \emph{Strong Janosik policy} or \emph{strong anti-Matthew policy}, $[b_+, b_-] = [2,1]$, with the increase in the equality preferred over the increase in the inequality, as in this case the reduction of the capital of the player with larger capital is bigger than the increase of his capital in the  case of negative outcome.

\item \emph{Strong Matthew policy}, $[b_+, b_-] = [-1,-2]$, with the increase in the inequality preferred over the decrease in the inequality, with the interpretation of the outcomes symmetric to the interpretations in the case of strong anti-Matthew policy.
\end{itemize}

In particular, in the case $b=[1,1]$, the procedure \gameref{proc:janosik-v2} is reduced to \gameref{proc:janosik-v1}. Thus, in the case of Janosik policy, the wining in the game is interpreted toward the reduction of inequality in the population.

%%%%%%%%%%%%%%%%%%%%%%%%%%%%%%%%%%%%%%%%%%%%%%%%%%%%%%%%%%%%%%%%%%%%%%%%%%%%%%%%
\section{Policy selection}\label{sec:policy-seclection}
%%%%%%%%%%%%%%%%%%%%%%%%%%%%%%%%%%%%%%%%%%%%%%%%%%%%%%%%%%%%%%%%%%%%%%%%%%%%%%%%

The behaviour of the scheme proposed in the previous section is controlled by the choice of the elementary games played by the agents. This choice can be crucial for the final results of the inequality reduction procedure. In this section we focus on two particular examples of elementary games used in the proposed scheme.

The most straightforward method of choosing the policy selection game is to use a bit flip game. In this case the policy is fair in the sense that on average only the difference of capitals will affect the exchange of capital between the players.

Another possible scenario is the selection policy governed by the Parrondo's scheme. In this case one can study the interplay between the personal average gain of the capital and the gain of the population expressed in terms of  inequality reduction.

In this section we provide the analysis of the numerical experiments conducted using the introduced schemes and using two proposed policy selection methods. The presented results were obtained using the following setup. Agents are distributed on a graph, in this case a regular 2D grid with $N\times N$ nodes, with periodic boundary conditions. The size of the grid is fixed to $N=10$. 
The graph is used to select agent which takes part in the elementary games. At each step, the agent is activated, and plays an elementary game against one of the agents occupying the same node.

In the presented numerical experiments, the activity of the agents is governed by \emph{random activation policy}~\cite{comer2014who}. According to this policy, each agent is activated once per step, and the activation is in random order.

For the purpose of the presented results we have utilized Mesa programming library~\cite{python-mesa-2020,mesa-github}. 
For the sake of reproducibility~\cite{turing-way,Leipzig2021} source code for the presented numerical experiments can be obtained from publicly accessible repository~\cite{matthew-repo}.

%%%%%%%%%%%%%%%%%%%%%%%%%%%%%%%%%%%%%%%%%%%%%%%%%%%%%%%%%%%%%%%%%%%%%%%%%%%%%%%%
\subsection{Random policy selection}\label{sec:basic-janosik-case}
%%%%%%%%%%%%%%%%%%%%%%%%%%%%%%%%%%%%%%%%%%%%%%%%%%%%%%%%%%%%%%%%%%%%%%%%%%%%%%%%

\begin{figure}[t!]
	\centering
	\includegraphics[scale=0.95]{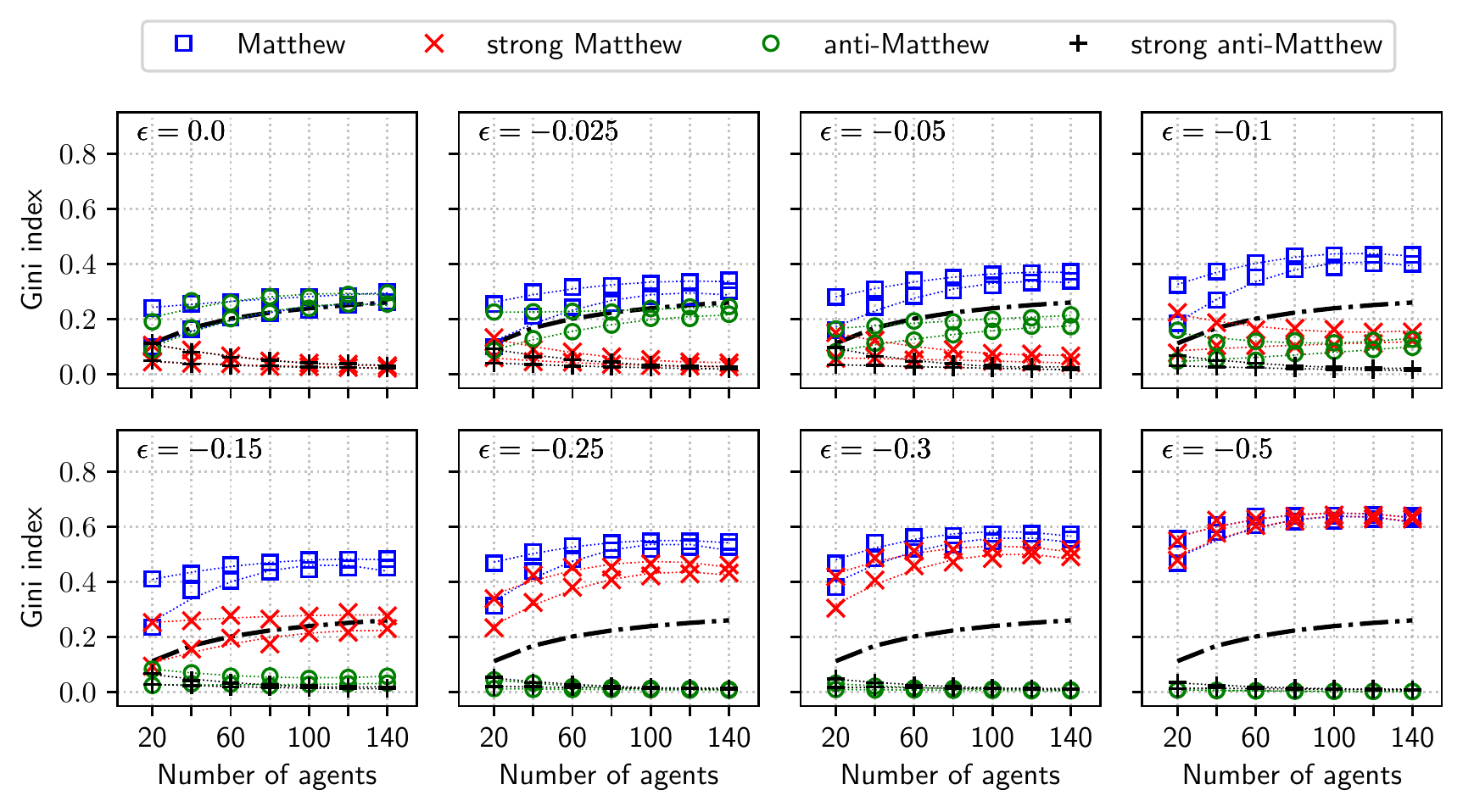}
	
	\caption{Inequality in the population for different policies and the bias parameter $\eps$ used in \gameref{proc:janosik-v2} for constant difference in the distribution of the initial capital. For each policy the maximal and the minimal values of the Gini index are marked.  Plots are based on the 250 steps of the procedure \gameref{proc:janosik-v2} repeated 50 times for each number of agents, $n=20,40,\dots,140$. Black dash-dotted line denotes the initial value of the Gini index for the initialization of capital given in Eq.~\eqref{eq:init-wealth-const}.}
	\label{fig:janosik-grid-10x10-50runs-250steps-uniform}
\end{figure}

\begin{figure}[t!]
	\centering
		\includegraphics[scale=0.95]{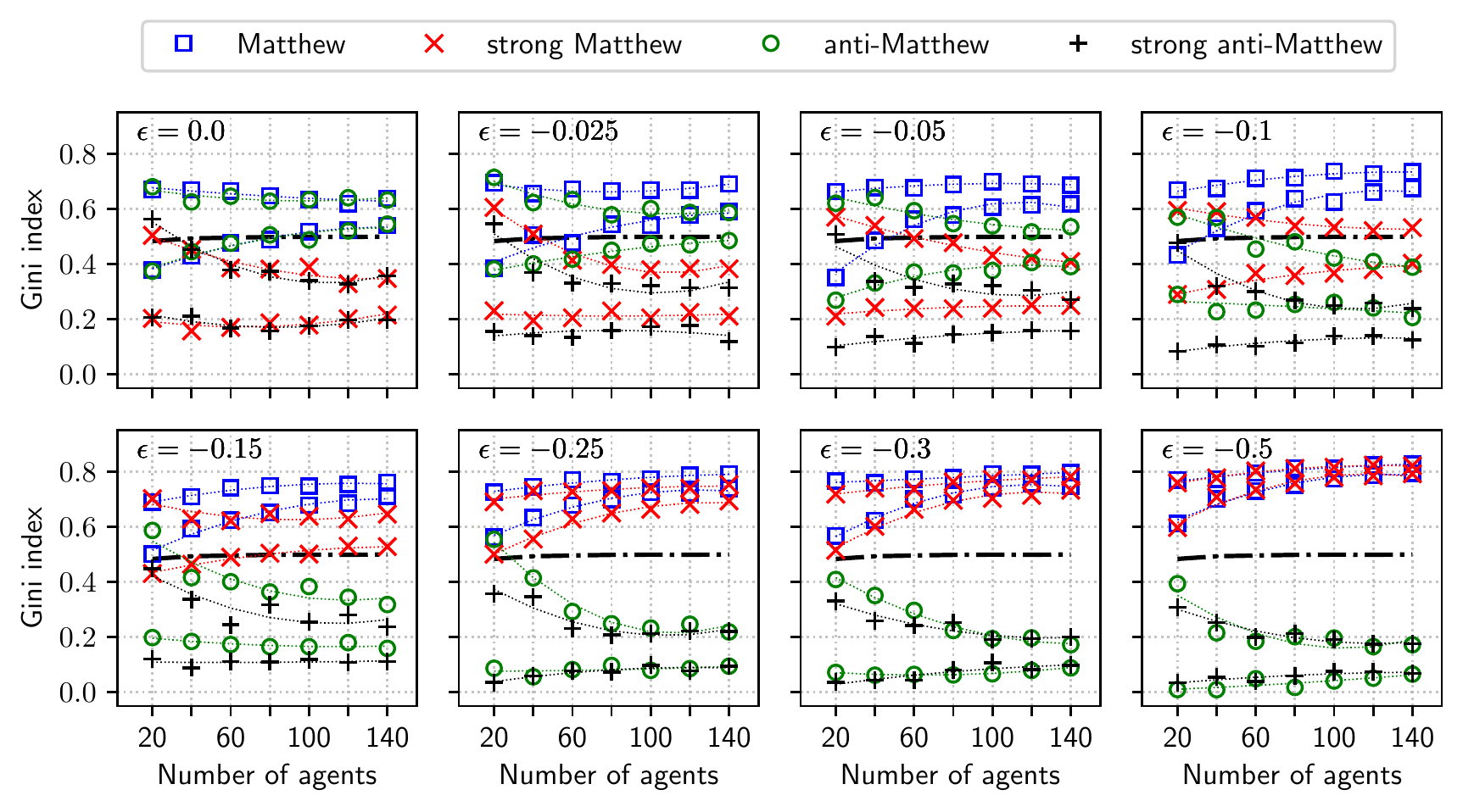}
	
	\caption{Inequality in the population for different policies and the bias parameter $\eps$ used in \gameref{proc:janosik-v2} with equilibrium distribution of the initial capital given by Eq.~\ref{eq:bg-dist}. Black dash-dotted line denotes the mean Gini index averaged over $10^4$ initial configurations. Total capital in the population is set to $20 n$, where $n$ is the number of agents. Other parameters are the same as in the case of Fig.~\ref{fig:janosik-grid-10x10-50runs-250steps-uniform}. }
	\label{fig:janosik-grid-10x10-50runs-250steps-boltzmann}
\end{figure}

In the simplest scenario the selection of policy is governed by the bit flip game. In this case one can consider the relation between the bias of the coin and the dynamics of the inequality. Numerical results obtained from the realization of the procedure described in \gameref{proc:janosik-v1} and \gameref{proc:janosik-v2} lead to the results presented in \figref{fig:janosik-grid-10x10-50runs-250steps-uniform}  and \figref{fig:janosik-grid-10x10-50runs-250steps-boltzmann}.

Parameter $\eps$ controls the bias of the game played by agents towards loosing, and its value corresponds to the average gain of the agent, which is given by $-2\eps$. In particular, value $\eps=0$ results in fair coin flip, and value $\eps=-0.5$ results in the situation where the agent always wins.

As one can observe in \figref{fig:janosik-grid-10x10-50runs-250steps-uniform}, for the case of a fair coin used during the elementary game played by the agents ($\eps=0$), both unbiased policies -- Matthew and anti-Matthew -- result in the preservation of the inequality. This is expected as this case is symmetric. Similarly, both biased policies -- strong Matthew and strong anti-Matthew -- lead to a significant reduction of the inequality in the capital distribution.

It is interesting to notice that even a relatively small change in the value of game bias parameter leads to a visible change in the effects of the unbiased policies. In the presented case, for $\eps=-0.025$, anti-Matthew (Janosik) policy leads to the desired goal and the final value of the Gini index is decreased.

One should also note that strong Matthew policy leads to the decrease of the inequality for the values of the game bias parameter $\eps\in\{0,-0.025,-0.05,-0.1,-0.15\}$. Only for $\varepsilon>-0.5$, the strong Matthew policy leads to the increase of the inequality. In this situation one can make the following observations:
\begin{itemize}
\item Anti-Matthew (Janosik) policy is not effective for the case $\eps=0$, which means that decrease in the Gini index is compensated by its increase. However, the application of strong anti-Matthew (strong Janosik) policy leads to the expected reduction of the Gini index.

\item The effect of Matthew policy is the increase of inequality, and the effect of anti-Matthew policy in the decrease of inequality is visible even for small non-zero values of $\eps$.

\item Surprisingly, for the case $\eps=0$, strong Matthew policy leads to the elimination of inequalities. The decrease of inequality obtained as the effect of strong-Matthew policy is visible for values up to $\eps=-0.125$.
\end{itemize}

Results of the numerical simulations for the case of equilibrium initialization of capital distribution are provided in \figref{fig:janosik-grid-10x10-50runs-250steps-boltzmann}. One can observer that the initial distribution has a significant impact on the behaviour of all strategies. 

In particular, in the case of equilibrium initialization of the capital, one can see that strong-Matthew policy is more effective. Already for  $\eps=-0.1$  such policy can lead to the increase in the inequality. Also in the symmetric interpretation expressed as Matthew policy, one can observe the increase in the inequality even for the unbiased case. This effect is more prominent that in the case of uniform difference in the initial capitals (cf. \figref{fig:janosik-grid-10x10-50runs-250steps-uniform}).

%%%%%%%%%%%%%%%%%%%%%%%%%%%%%%%%%%%%%%%%%%%%%%%%%%%%%%%%%%%%%%%%%%%%%%%%%%%%%%%%
\subsection{Parrondo's scheme policy selection}\label{sec:parrondo-case}
%%%%%%%%%%%%%%%%%%%%%%%%%%%%%%%%%%%%%%%%%%%%%%%%%%%%%%%%%%%%%%%%%%%%%%%%%%%%%%%%

In the schemes considered in procedures \gameref{proc:janosik-v1} and \gameref{proc:janosik-v2} we assumed that the inequality regulation policy is applied using a simple coin flip game. This simplifies the model, but it is interesting to see how does the form of the elementary game influences the values of capitals.

As an alternative example of the elementary game played by the agents we use Parrondo's scheme. In this case we consider two elementary game, game \gameA and game \gameB. The choice of the particular game used at each step depends on the strategy of the player. Additionally, the average gain for the selected game can depend on the current capital possessed by the player. Such construction leads to the procedure described in \gameref{proc:janosik-parrondo}.

\begin{algorithm}
\begin{algorithmic}
    \REQUIRE  games \gameA and \gameB defined by $\delta$ and $K$, game selection policy $\mathcal{P}$, graph $G$, $m_i \geq 0, i=1,2\dots, n$, and interpretation policy bias $b=[b_{+},b_{-}]$, 
    \FOR{$i=1$ to $n$}
        \STATE $j$ $\leftarrow$ choose another agent from the node occupied by agent $i$
        \STATE $r$ $\leftarrow$ result of the Parrondo's game played by agent $i$ using policy $\mathcal{P}$
        
       	\STATE $\Delta \leftarrow \sgn{(m_i-m_j)}$
      \IF{$r > 0$} 
        \STATE $(m_i, m_j) \leftarrow (m_i-b_+\Delta, m_j+b_+\Delta)$
      \ELSE
        \STATE $(m_i, m_j) \leftarrow (m_i+b_-\Delta, m_j-b_-\Delta)$ 
      \ENDIF
%    \IF{$r = 1$} 
%        \STATE $m_i \leftarrow m_i-b_{+}\sgn{(m_i-m_j)}$
%        \STATE $m_j \leftarrow m_j+b_{+}\sgn{(m_i-m_j)}$
%    \ELSIF{$r = -1$}
%        \STATE $m_i \leftarrow m_i+b_{-}\sgn{(m_i-m_j)}$
%        \STATE $m_j \leftarrow m_j-b_{-}\sgn{(m_i-m_j)}$
%    \ENDIF
%        \STATE {select one of the neighbour nodes for the current node occupied by $i$}
        \STATE {move agent $i$ to the new position}
    \ENDFOR
\end{algorithmic}
\caption{Scheme for the biased interpretation of the results with policy bias and policy for choosing games according to the Parrondo's scheme.}
\label{proc:janosik-parrondo}
\end{algorithm}

In this scenario each step consists of the following actions.
\begin{itemize}

\item If there are any other agents at the current node, choose a game which will be played during the step. The choice of the played game depends on the game selection policy.

\item Play the selected game from the Parrondo's scheme using the value of the capital of the agent.

\item Alter the value of the capital for both players depending on the outcome of the game and taking into account the inequality controlling policy.
\end{itemize}

Values of the bias parameter used in the Parrondo's scheme was selected from to $\delta = 0.005$ to $\delta = 0.1$. The parameter governing the behaviour of game $\gameB$ was set to $K=3$. Combination $\delta = 0.005$ and $K=3$ leads to the example of Parrondo's paradox in the capital dependant scenario for using randomly selected game~\cite{harmer2002review,gawron2005quantum}.

\begin{figure}[t!]
\includegraphics[scale=0.95]{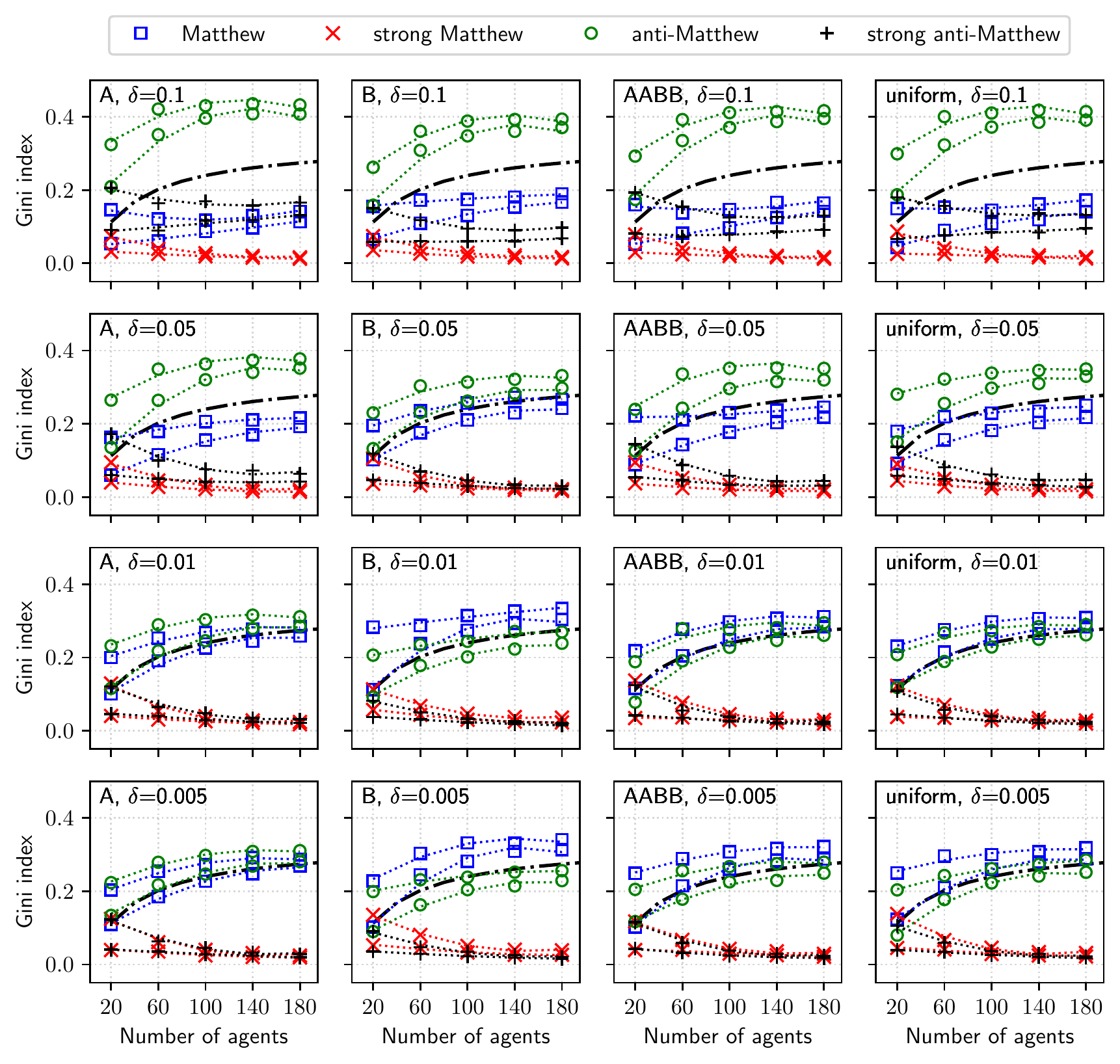}

\caption{Influence of the parameters used in the Parrondo's scheme in \gameref{proc:janosik-parrondo} on the changes in the Gini index. In this case four policies for constructing Parrondo's scheme were used, namely: always play game \gameA, always play game \gameB, play games \gameA and \gameB in sequence \game{AABB}, and choose which game to play randomly with equal probability. Parameter $\delta\in\{0.1,0.05, 0.01, 0.005\}$ is used in games \gameA and \gameB used in the Parrondo' scheme. Plots are based on the 250 steps of the procedure \gameref{proc:janosik-v2} repeated 50 times for each number of agents, $n=20,60,\dots,180$. Black dash-dotted line denotes the initial value of the Gini index for the initialization of capital with constant difference (cf. Eq.~\eqref{eq:init-wealth-const}).}
\label{fig:janosik-grid-10x10-50runs-250steps-parrondo}
\end{figure}

The results for the numerical experiments conducted in the case of  \gameref{proc:janosik-parrondo} are presented in \figref{fig:janosik-grid-10x10-50runs-250steps-parrondo}. From the obtained results one can conclude that anti-Matthew policy, as well as Matthew policy, are not effective in this case. They do not lead to the expected results for the value of parameter $\delta>0.05$. Surprisingly, strong Matthew policy is almost not affected by changes in the game parameters. Moreover, it leads to the significant decrease in the value of the Gini index in all considered cases. Additionally, anti-Matthew policy is effective only for $\delta<0.01$. It can be observed that for $\delta=0.005$ and the uniform selection of elementary games in the Parrondo's scheme, the anti-Matthew policy leads to the expected reduction of inequality in the capital distribution. However, as this effect is visible also for the  \gameB and \game{AABB} selection of elementary games, it cannot be attributed to the paradoxical effect which is observed in the case of the uniform selection of elementary games in the Parrondo's scheme. On the contrary, for $\delta=0.01$, and the elementary game \game{AABB}, the effect of inequality reduction is less prominent that in the case of using \gameB only. This suggest that the paradoxical effect observed in the Parrondo's scheme in terms of average gain, is not observed in terms of inequality reduction. Moreover, using losing game only -- in this case game \gameB\ -- can be beneficial from the point of view of inequality reduction.

%%%%%%%%%%%%%%%%%%%%%%%%%%%%%%%%%%%%%%%%%%%%%%%%%%%%%%%%%%%%%%%%%%%%%%%%%%%%%%%%
\section{Final remarks}\label{sec:discussion}
%%%%%%%%%%%%%%%%%%%%%%%%%%%%%%%%%%%%%%%%%%%%%%%%%%%%%%%%%%%%%%%%%%%%%%%%%%%%%%%%

The goal of this work was to introduce a scheme for reducing the inequality in the capital distribution among the agents on the network. To this end we proposed an interpretation of the game on network in terms of inequality reduction, and we considered symmetric and asymmetric interpretation policies.  

We also studied two particular examples for controlling the elementary game selection. In the first scenario, we used a biased coin to mimic the random selection of the policy for reducing the inequality. We demonstrated how the value of the bias influences the resulting Gini index.

In the second scenario, we investigated the relation between the Parrondo's effect and the capital inequality control. Following the scheme used in the Parrondo's paradox, we provided a construction of a scheme composed of two losing games. We used the result of the game to introduce the behaviour of suppressing the inequality of wealth distribution in the group of agents in a network. We also provided arguments that the interpretation of the Parrondo paradox in terms of the capital cannot be translated into the interpretation in terms of the capital reduction.

One can note that the schemes studied in the presented work do not include mechanics taking into account the information about the global inequality. In the presence of such global governing mechanism  such scenarios, the information about the value of the inequality index could be incorporated into the inequality regulation policy. In the schemes introduced in the presented paper, the agents were not aware of the global inequalities in the capital distribution. The interpretation of the local game was based on the difference between the players only. It is natural to ask if the global mechanics for distributing information about the inequalities, expressed by the Gini index or by other inequality indicators, can lead to a different behaviour of the model.

Another possible extension of the presented work is the study of the proposed schemes in the case where the interactions between agents take place on the complex network. Such extension could be useful to study the possibility of managing the wealth interpreted not as capital, but as other forms of social factors.

It is also worth stressing that the presented study is limited by the applicability of the Gibbs distribution. In scenarios observed in economy and social science one cannot assume that the system is in the equilibrium. In such systems it is necessary to take into account the permanent dynamics of the processes governing the population.

Finally, other possible methods of capital initialization could be investigated. In particular, one could mimic the wealth distribution obtained in some communities or countries, and thus test the proposed model in the real-world context. Such study could be particularity interesting if the historical data and the past decisions could be described and analysed in the context of social inequality reduction.

%%%%%%%%%%%%%%%%%%%%%%%%%%%%%%%%%%%%%%%%%%%%%%%%%%%%%%%%%%%%%%%%%%%%%%%%%%%%%%%%
\section*{Acknowledgments}
%%%%%%%%%%%%%%%%%%%%%%%%%%%%%%%%%%%%%%%%%%%%%%%%%%%%%%%%%%%%%%%%%%%%%%%%%%%%%%%%

Author would like to thank Nenggang Xie, Ye Ye and Wang Meng for motivating the presented study and for their hospitality during the visit at the Anhui University of Technology.

%%%%%%%%%%%%%%%%%%%%%%%%%%%%%%%%%%%%%%%%%%%%%%%%%%%%%%%%%%%%%%%%%%%%%%%%%%%%%%%%
%\bibliographystyle{elsarticle-num}
%%\bibliography{matthew_reduction_game}
%%%%%%%%%%%%%%%%%%%%%%%%%%%%%%%%%%%%%%%%%%%%%%%%%%%%%%%%%%%%%%%%%%%%%%%%%%%%%%%%

\end{document}